\documentclass{vgtc}                          

\graphicspath{{figures/}{pictures/}{images/}{./}} 

\usepackage{times}                     

\usepackage{tabu}                      
\usepackage{booktabs}                  
\usepackage{lipsum}                    
\usepackage{mwe}                       

\usepackage{mathptmx}                  

\onlineid{0}

\vgtccategory{Research}

\vgtcinsertpkg

\PassOptionsToPackage{backref=false}{hyperref}
\PassOptionsToPackage{naturalnames}{hyperref}

\title{Fostering Creative Visualisation Skills Through Data-Art Exhibitions}

\author{Jonathan C. Roberts\thanks{e-mail: j.c.roberts@bangor.ac.uk}\\%
        \scriptsize Bangor University %
}

\abstract{
Data-art exhibitions offer a unique and real-world setting to foster creative visualisation skills among students. They serve as real-world platform for students to display their work, bridging the gap between classroom learning and professional practice. Students must develop a technical solution, grasp the context, and produce work that is appropriate for public presentation. This scenario helps to encourage innovative thinking, engagement with the topic, and helps to enhance technical proficiency. We present our implementation of a data-art exhibition within a computing curriculum, for third-year degree-level students. Students create art-based visualisations from selected datasets and present their work in a public exhibition. We have used
this initiative over the course of two academic years with different cohorts, and reflect on its impact on student learning and creativity.
} 


\renewcommand*{\backref}[1]{
}

\begin{document}

\firstsection{Introduction}
\maketitle
Data art merges several disciplines, including data science, visualisation and art. The aim is to transform raw data into visual narratives. While visualisation encourages accurate and correct information display, the emphasis on data art is on creative expression. In other words, the focus shifts away from precision and exactness aimed at qualitative understanding, to delivering emotional responses, stimulation of thought, and creation of visually captivating and intriguing artworks. 
The creation of data art offers a valuable project scenario for students. It constitutes an \textit{authentic}~\cite{Swaffield2011_HeartOfAuthenticAssessment} task as it necessitates that students grasp real-world issues, overcome technical challenges, and produce work that is accessible to the public.
An authentic task is a learning activity that involves real-world challenges, requiring students to apply their skills and knowledge in a meaningful context. It bridges the gap between theoretical classroom learning and real-life application, helping students develop skills directly relevant to their future careers or personal lives.

We run a Creative Visualisation course, that is an option for third-year computing students. The students are following their BSc in Computer Science or Creative Technology. The course is 20 credits, which corresponds to 200 hours of student effort, including contact hours (lectures and classes), and independent study (reading, research and assignments). As part of the students' assessment, we require them to produce a data art piece based on a dataset of their choice. This piece is intended to be visually compelling and conceptually insightful, transforming raw data into an engaging and meaningful artwork.

\begin{figure*}
    \centering
    \includegraphics[width=\textwidth]{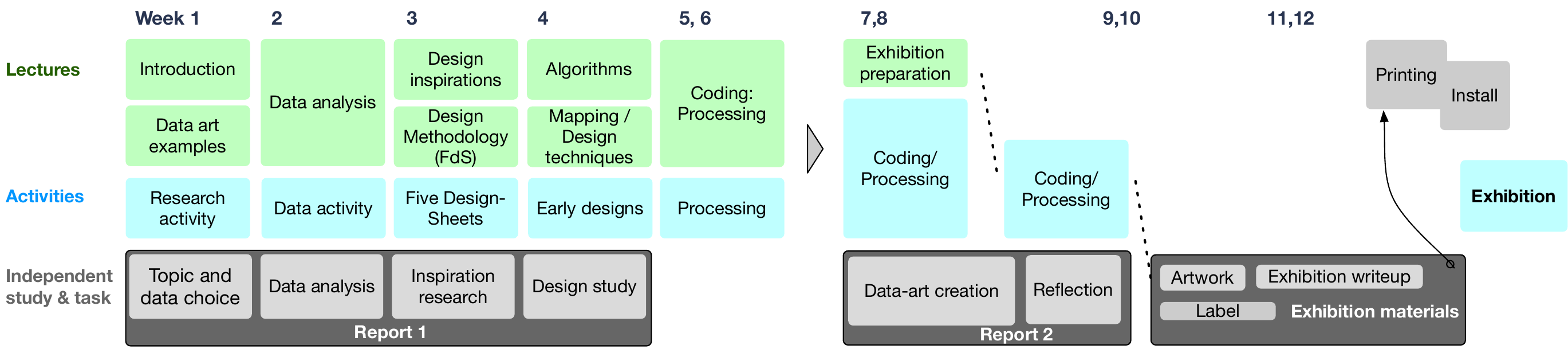}
    \caption{The schedule of lectures, activities and independent study, for the data-art exhibition authentic task. This project-based learning enables students to focus on a task, learn about data analysis, design, generate inspired ideas, and implement a data-art artwork that is suitable for public display.}
    \label{fig:schedule}
\end{figure*} 

We explain the process, describe the 12 week activities, assessments and exhibition, and discuss its benefits and challenges, such that others can follow.

\section{Related work}
Fostering creativity in visualisation, especially for students and learners can be a huge challenge, and there are many often conflicting goals for an educator to balance. ``\textit{Instructors need to design teaching situations, motivate students to learn, actively promote the topic and its purpose ..}''~\cite{BachEtal2024}.
We inspire students through the exhibition, an authentic task that engages them in real-world challenges by requiring them to create and publicly display data art that reflects thoughtful consideration of the situation.
In an educational context, creative tasks must strike a balance: they should be open-ended to foster creativity yet defined enough for effective grading and to provide focus for students. The task of creating data art for exhibition accomplishes these objectives by addressing the fundamental questions: who, what, why, when, how, and where. This clarity is crucial as it helps students stay focused, provides them with a clear goal, and allows teachers to clearly outline task requirements.

One of the first tasks in the course is to define data art for the students. This is done by presenting various examples and referencing the definition by Viegas and Wattenberg~\cite{ViegasWattenberg2007}, who describe data art as ``\textit{visualisations of data created by artists with the intention of producing art}''. The intent behind the output is crucial. The approach, intent, and audience for data art differ significantly from those of typical visualisations. Data art seeks to evoke emotions, convey a narrative, and emphasise the beauty of design, prioritising aesthetic appeal over scientific quantitative value~\cite{kosara2007visualization}. The medium and presentation method of data-art also differs, from pen, paper, ink, paint to physical objects. In our lectures we explain many examples, including sketches of `dear data'~\cite{lupi2016dear}, glyphs of dataQuilt~\cite{zhang2020dataquilt}, video of  \"Angeslev\"a and Cooper~\cite{angeslevaCooper}, photographs of Paolo Cirio (\href{https://www.paolocirio.net}{paolocirio.net}), and paper and rope weather visualisations by Nathalie Miebach (\href{https://www.nathaliemiebach.com}{nathaliemiebach.com}).

\section{Lectures, activities and study tasks}
The teaching takes place over 12 weeks, and each week they have lectures, in-class activities and tasks to do for  independent study. Fig.~\ref{fig:schedule} presents the schedule. Each week includes a \textbf{lecture} and an \textbf{activity}, with students assigned a \textbf{task} to complete independently. The work splits into three stages. They analyse the data, research and perform a design study, that they submit at the end of the 6 weeks period as report~1. During the next six weeks the students implement their design, and submit the reflective report (report~2).
Finally they submit their artwork, exhibition write-up and 120 word label. The artwork is printed and then installed.

Students select a topic and corresponding dataset, which is registered with the teacher by the end of week~1. If they cannot find a suitable dataset, one is provided from a tutor-curated collection. The independent study tasks dictate the organisation and sequencing of the taught content. For instance, to complete the data analysis task, students receive a lecture on data analysis along with an activity focused on the topic. The first activity is about finding data-art examples and their own topic. Week~2 centers on data analysis, with students tasked to deconstruct the data into its various components (categorical, continuous, etc.). Week~3 focuses on design methodology. We start by lecturing about different art styles, encouraging students to recognise the vast diversity within art. From abstract interpretations and detailed realism to avant-garde techniques, impressionistic approaches, and contemporary trends in artistry, we cover a spectrum of artistic styles. This broad exploration not only fosters an appreciation for the breadth of artistic expression but also inspires students to integrate these diverse styles into their own creative processes.
We teach the Five Design-Sheets (FdS)~\cite{RobertsHeadleandRitsos16}
method, and give an activity on sketching alternatives, and sketching using the FdS method. We also provide numerous examples of data-art (refer to Related Work for some examples). Students present their design-study and FdS in their report. Week 4 delves into algorithms and mapping techniques. We explore algorithms such as pixel bar charts~\cite{KeimPixelBarCharts2002}, Diffusion Limited Aggregation (DLA), particle systems, L-systems, Genetic Algorithms, and more. By this time, the students can write and submit their first report, which summarises their analysis, research inspiration and design study.

The second part focuses on creating the data-art.
We lecture on the practicalities of exhibition preparation and coding, and run a variety of activities focused on processing, helping students practice tasks such as plotting, layout design, loading data, saving to PDF, and more. Students use \href{http://processing.org}{processing.org} to create their data art, and we provide continuous feedback on their progress. 
We use Processing to encourage creative design, it is easy to create PDF vector output, and it leverages students' prior experience with Java. 
They present their data art creation in a reflective report.
Third, they deliver their exhibition artwork (as a PDF), and include a 120 word label, and exhibit commentary (of 500 words). The commentary is included in the exhibition additional materials; the artwork and label are printed professionally. This also means that students must submit their work on time for the print run.

\section{Discussion and Conclusions}
The data-art exhibition represented the finale of the students' academic journey. It provided them with the opportunity to analyse datasets, explore various artistic styles, and produce data-art pieces that were showcased publicly for their peers to see. 
One student shared that they were ``\textit{delighted and excited to be a part of the exhibition}'', while another expressed pride in seeing their work on display, saying, ``\textit{this is my work}''. It achieved the intended outcome of sparking students' excitement and aiding them in crafting a high-quality submission.
The reports clearly demonstrated that the students had to understand a variety of challenges (and especially practical issues of data visualisation/data art). From understanding the dimensions of data, performing a data analysis, cleaning the data, removing missing parts, organising the narrative, designing suitable solutions, mapping the data, creating colourmaps, and so on. They needed to consider how their work would be suitable for a public audience, offering valuable real-world scenarios. 

The exhibition illustrated the diversity and scope of data-art. Some students did opt for traditional visualisation methods (there were a few who based their visualisations on scatter plots, and bar charts). The majority explored innovative approaches. But each of the students moved away from conventional dashboards, and some were extremely creative. The activities in the class also helped students to share their ideas. They could help each other, without challenges of plagiarism and copying.

\begin{figure}[t]
    \centering
    \includegraphics[width=\columnwidth]{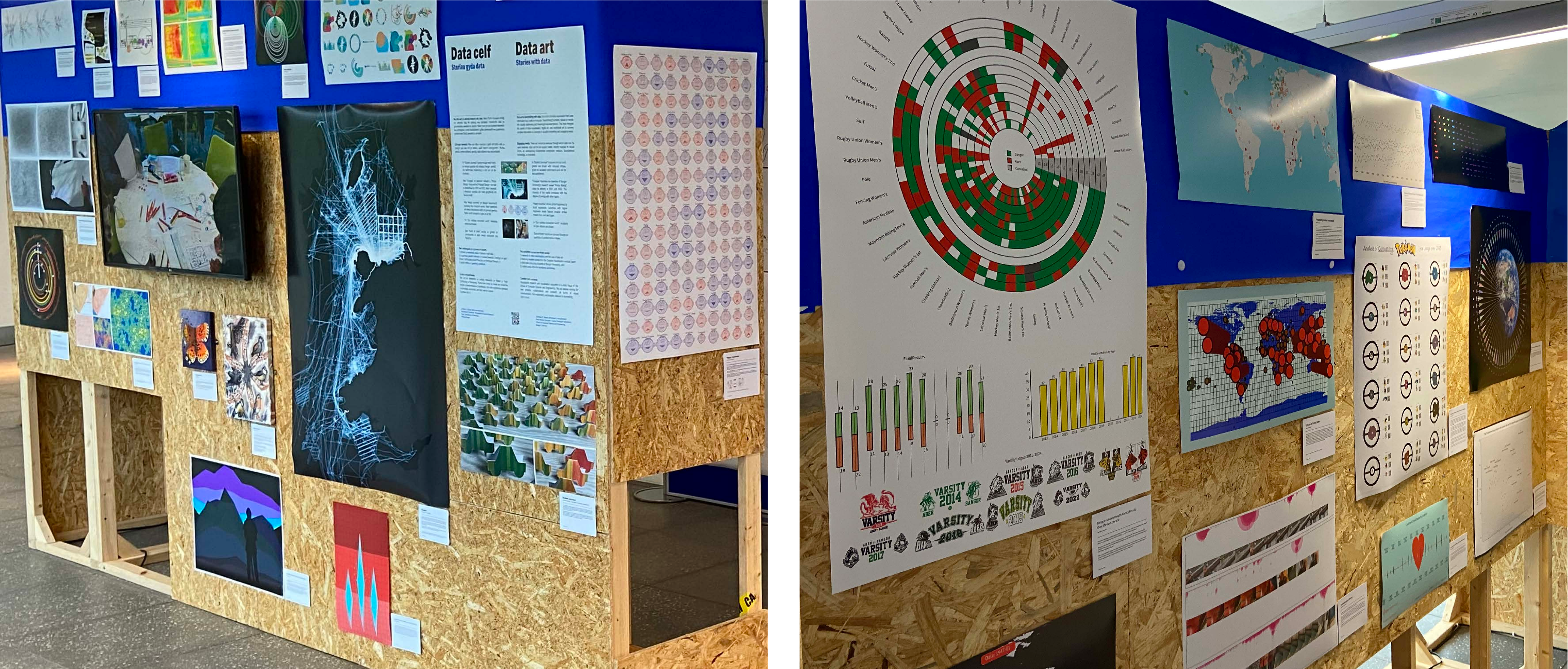}
    \caption{Final exhibition showing students' work. Featuring a diverse array of data-art creations crafted by students as part of their coursework.\vspace{-2mm}}
    \label{fig:enter-label}
\end{figure}

\bibliographystyle{abbrv-doi-hyperref-narrow}

\bibliography{fosteringCreativeVisSkills}
\end{document}